\documentclass[12pt]{article}

\usepackage{graphicx}

\begin{document}

\title{Energy Level Statistics of the U(5) and O(6)
 Symmetries in the Interacting Boson Model}

\author{{ Jing Shu$^{1}$, Ying Ran$^{1}$, Tao Ji$^{1}$, Yu-xin Liu$^{1,2,3,
4,}$\thanks{Corresponding author} }\\[3mm]
\normalsize{$^1$ Department of Physics, Peking University, Beijing 100871, China}\\
\normalsize{$^2$ The Key Laboratory of Heavy Ion Physics of the
Chinese Ministry of Education, }\\
\normalsize{Peking University, Beijing 100871, China}\\
\normalsize{$^3$ Institute of Theoretical Physics, Academia Sinica,
Beijing 100080, China} \\
\normalsize{$^4$ Center of Theoretical Nuclear Physics, National
Laboratory of}\\ \normalsize{ Heavy Ion Accelerator, Lanzhou
730000, China} }

\maketitle

\begin{abstract}
 We study the energy level statistics of the states in U(5) and O(6)
dynamical symmetries of the interacting boson model and the high
spin states with backbending in U(5) symmetry. In the
calculations, the degeneracy resulting from the additional quantum
number is eliminated manually. The calculated results indicate
that the finite boson number $N$ effect is prominent. When $N$ has
a value close to a realistic one, increasing the interaction
strength of subgroup O(5) makes the statistics vary from
Poisson-type to GOE-type and further recover to Poisson-type.
However, in the case of $N \rightarrow \infty$, they all tend to
be Poisson-type. The fluctuation property of the energy levels
with backbending in high spin states in U(5) symmetry involves a
signal of shape phase transition between spherical vibration and
axial rotation.
\end{abstract}

\bigskip


\newpage

\parindent=20pt

\baselineskip=24pt

\section{INTRODUCTION}
Random-matrices theory (RMT)\cite{meta67} provides a basis to
study quantum chaotic systems. Particularly, the fluctuation
properties of fully chaotic systems with time reversal symmetry
follow the Gaussian orthogonal ensemble (GOE) whereas nonchaotic
ones follow Poisson ensemble\cite{Brod81}. Notice that dynamical
symmetry means the integrability of the system in classical limit
and constants of motion associated with a symmetry govern the
integrability of the system, investigating the effects of symmetry
is of importance to study the dynamics of a quantum system. In
recent years, many numerical studies concerning different types of
symmetries and their relations to the onset of chaos have been
carried out\cite{Zhang88, MKZ88,PV90,PVD92,WAL93, LKC94, LW96},
however the case that different types of symmetries coexist and
compete with each other in one quantum system has not yet been
analyzed carefully. In this point of view, we study the nucleus in
certain dynamical symmetries which were expected to be completely
integrable in the past.

It has been known that the interacting boson model
(IBM)\cite{IA87} is a realistic theoretical model in describing
the low-energy collective states and the electromagnetic
transitions of a large number of even-even nuclei successfully. In
the original version of the IBM (IBM1), nuclei are regarded as
systems composed of $s$- and $d$-bosons with symmetry U(6), and it
has three dynamical symmetries U(5), SU(3) and O(6), geometrically
corresponding to spherical vibration, axial rotation and
$\gamma$-unstable rotation\cite{GK80}, respectively. Because
collective motion is described by a hamiltonian matrix of finite
dimension, one can diagonalize the matrix easily and study the
energy level statistics such as nearest-neighbor spacing
distribution(NSD) numerically to check whether the motion is
chaotic or regular. Then there have been many works to investigate
the fluctuations of the nucleus by analyzing the energy level
statistics (see, for example, Refs.\cite{AN92,AW91,ANW90}) in the
framework of the IBM. However, except for the case of SU(3)
symmetry, the energy level statistics has not yet been analyzed
for dynamical symmetries. Such a neglect is quite natural since,
according to the symmetry paradigm, the energy level statistics
should be Poisson-type. Nevertheless, the investigation on the
SU(3) symmetry showed that the statistics depended strongly on the
boson number of the system $N$ and it was quite close to GOE
statistics in the realistic cases where $N$ was not very
large.\cite{PVD92}. In this aspect, the energy level statistics of
the states in a dynamical symmetry may be more complicated than
the symmetry paradigm predicts. We will then analyze the energy
level statistics of the U(5) and O(6) symmetries in this work. For
comparison, we also involve the SU(3) limit.

More recently, a breakthrough has been carried out by Iachello in
the study of critical point behavior of the nucleus undergoing a
shape-phase transition. It has been shown that the critical point
of the transition between vibration and $\gamma$-unstable rotation
and that between vibration and axial rotation hold the symmetry
E(5), X(5)\cite{Iac00,Iac01}, respectively. Although fluctuation
properties of these transitional regions have been studied by
Alhassid and collaborates\cite{AN92,AW91,ANW90}, the statistics at
the critical points has not been discussed in detail. On the other
hand, investigating the property of high spin nuclear states and
the mechanism of backbending of high spin states has long been a
significant topic in nuclear physics. It has been known that the
backbending comes from the breaking of nucleon pairs and the
alignment of the angular momenta. Recently, another way for the
backbending, more concretely, the collective backbending to appear
has been proposed to be a property of the U(5) symmetry of the
IBM\cite{Long97}. In such a formalism, with a special way to fix
the parameters, the yrast states with the U(5) symmetry change
from the vibrational ones with different $d$-boson numbers to the
rotational ones with full d-boson configuration($n_{d}=N$) when
the angular momentum $L$ reaches a critical value $L_c$. In this
sense, the energy level structure of the states in the U(5)
symmetry might have a sign of shape phase transition. We then
analyze the statistics as the first step to explore the
fluctuations of a shape phase transition system.

The paper is organized as follows. In Section 2, we survey the
framework of the IBM and the method to analyze the energy level
statistics briefly. In Section 3, we represent the numerical
results and give some discussions. Finally, a summary and some
remarks are given in Section 4.

\section{METHOD}

In the original version of the IBM (IBM1), the collective states
of nuclei are described by $s$- and $d$-bosons. The corresponding
dynamical group is U(6), and it has three dynamical symmetry
limits U(5), O(6) and SU(3). Taking into account one- and two-body
interactions among the bosons, one has the Hamiltonian of the
nucleus with one of the three dynamical symmetries as\cite{IA87}
$$\displaylines{\hspace*{1cm}
H_{U(5)}=E_0 + \varepsilon C_{1U(5)} + \alpha C_{2U(5)} +\beta
C_{2O(5)}+ \gamma C_{2O(3)} \, , \hfill{(1)} \cr }
$$
$$\displaylines{\hspace*{1cm}
H_{O(6)}=E_0 + \eta C_{2O(6)} + \beta C_{2O(5)} + \gamma C_{2O(3)}
\, , \hfill{(2)} \cr }
$$
$$\displaylines{\hspace*{1cm}
H_{SU(3)}=E_0 + \delta C_{2SU(3)} + \gamma C_{2O(3)} \, .
\hfill{(3)} \cr }
$$
In case of the U(5), O(6) or SU(3) symmetry, the wave-function can
be expressed as
$$\displaylines{\hspace*{1cm}
|\psi_{U(5)}\rangle=|N n_d \tau K L \rangle \, , \hfill{(4)}\cr }
$$
$$\displaylines{\hspace*{1cm}
|\psi_{O(6)}\rangle=|N \sigma \tau K L \rangle \, , \hfill{(5)}\cr
}
$$
$$\displaylines{\hspace*{1cm}
|\psi_{SU(3)}\rangle=|N (\lambda, \mu) K L \rangle \, ,
\hfill{(6)}\cr}
$$
where $N$ is the total number of the bosons, $n_d, \sigma,
(\lambda, \mu), \tau,
 L$ are the irreducible representations(IRREPs) of the group
U(5), O(6), SU(3), O(5) and O(3), respectively. $K$ is the
additional quantum number to distinguish the degenerate states
which have the same quantum number of the parent group.

If a nucleus is in one of the above mentioned dynamical
symmetries, its energy can be given by the IRREPs as
$$\displaylines{\hspace*{1cm}
E_{U(5)}=E_0 + \varepsilon n_d + \alpha n_d(n_{d}+4) +\beta
\tau(\tau+3)+ \gamma L(L+1) \, , \hfill{(7)} \cr }
$$
$$\displaylines{\hspace*{1cm}
E_{O(6)}=E_0 + \eta \sigma(\sigma+4) + \beta \tau(\tau+3) + \gamma
L(L+1) \, , \hfill{(8)} \cr }
$$
$$\displaylines{\hspace*{1cm}
E_{SU(3)}=E_0 + \delta
(\lambda^{2}+\mu^{2}+\lambda\mu+3\lambda+3\mu) + \gamma L(L+1) \,
. \hfill{(9)} \cr }
$$

To analyze the energy level statistics of the states in the
dynamical symmetries, we take the following process. At first,
with Eqs.(7), (8) and (9), we calculate the energy levels of the
nucleus in U(5), O(6) or SU(3) symmetry in IBM with different
total boson number $N$, spin-parity $J^\pi$ and several sets of
parameters $\alpha, \beta, \gamma, \delta, \varepsilon, \eta$.

For a given spectrum $\{E_i\}$, it is necessary to separate it
into the fluctuation part and the smoothed average part whose
behavior is nonuniversal and can not be described by random-matrix
theory(RMT)\cite{meta67}. To do so we take the unfolding process
for the energy spectrum (see for example Ref.\cite{AN92}). At
first we count the number of the levels below $E$ and write it as
$$\displaylines{\hspace*{1cm}
N(E)=N_{av}(E)+N_{fluct}(E) \, . \hfill{(10)} \cr }
$$
Then we fix the $N_{av}(E_i)$ semiclassically by taking a smooth
polynomial function of degree 6 to fit the staircase function
$N(E)$. We obtain finally the unfolded spectrum with the mapping
$$\displaylines{\hspace*{1cm}
\{\widetilde{E}_i \}=N(E_i) \, . \hfill{(11)} \cr }
$$
This unfolded level sequence $\{\widetilde{E}_i\}$ is obviously
dimensionless and has a constant average spacing of 1, but the
actual spacings exhibit frequently strong fluctuation.

We have used two statistical measures to determine the fluctuation
properties of the unfolded levels: the nearest neighbor level
spacings distribution(NSD) $P(S)$ and the spectral rigidity
$\Delta_3(L)$. The nearest neighbor level spacing is defined as
$S_i=(\widetilde{E}_{i+1})-(\widetilde{E}_i)$. The distribution
$P(S)$ is defined as that $P(S)dS$ is the probability for the
$S_i$ to lie within the infinitesimal interval $[S, S+dS]$. It has
been shown that the nearest neighbor spacing distribution $P(S)$
measures the level repulsion (the tendency of levels to avoid
clustering) and short-range correlations between levels. For a
regular system, it is expected to behave like the Poisson
statistics
$$\displaylines{\hspace*{1cm}
P(S)=e^{-S} \, , \hfill{(12)} \cr }
$$
whereas if the system is chaotic, one expects to obtain the Wigner
distribution
$$\displaylines{\hspace*{1cm}
P(S)=(\pi/2)S \ \textrm{exp}(-{\pi}S^{2}/4) \, , \hfill{(13)} \cr
}
$$
which is consistent with the GOE
statistics\cite{meta67,Brod81,port65}. With the Brody parameter
$\omega$ in the Brody distribution
$$\displaylines{\hspace*{1cm}
P_{\omega}(S)=\alpha(1+\omega)S^{\omega}
\textrm{exp}({-\alpha}S^{(1+\omega)}) \, , \hfill{(14)} \cr }
$$
where
$$\displaylines{\hspace*{1cm}
\alpha=\Gamma[(2+\omega)/(1+\omega)]^{1/2} \, \hfill{(15)} \cr }
$$
and $\Gamma[x]$ is the $\Gamma$ function, the transition from
regularity to chaos can be measured with the Brody parameters
$\omega$ quantitatively. It is evident that $\omega=1$ corresponds
to the GOE distribution, while $\omega=0$ to the Poisson-type
distribution. A value $0<\omega<1$ means an interplay between the
regular and the chaotic.

As to the spectral rigidity $\Delta_{3}(L)$, it is defined as
$$\displaylines{\hspace*{1cm}
\Delta_{3}(L)={\Bigg\langle}min_{A,B}\frac{1}{L}\int_{-L/2}^{L/2}
[N(x)-Ax-B]^{2}dx{\Bigg\rangle} \, , \hfill{(16)} \cr }
$$
where $N(x)$ is the staircase function of a unfolded spectrum in
the interval $[-L/2,x]$. The minimum is taken with respect to the
parameters $A$ and $B$. The average denoted by
$\langle\cdots\rangle$ is taken over a suitable energy interval
over $x$. Thus from this definition $\Delta_{3}(L)$ is the local
average least square deviation of the staircase function $N(x)$
from the best fitting straight line. It has also been shown that
the spectral rigidity $\Delta_{3}(L)$ signifies the long-range
correlation of quantum spectra\cite{Brod81} which make it possible
that for a chaotic spectrum very small fluctuation of the
staircase function around its average can be found in an interval
of given length (the interval may cover dozens of level spacings).
For the GOE the expected value of $\Delta_{3}(L)$ can only be
evaluated numerically, but it approaches the value
$$\displaylines{\hspace*{1cm}
\Delta_{3}(L)\cong(lnL-0.0687)/\pi^2 \, \hfill{(17)} \cr }
$$
for large $L$.  and for Poisson statistics
$$\displaylines{\hspace*{1cm}
\Delta_{3}(L)=L/15 \, . \hfill{(18)} \cr }
$$

\section{NUMERICAL RESULTS AND DISCUSSION}

At first, we analyze the energy levels given in Eqs.(7) and (8)
for the U(5) and O(6) dynamical symmetries, respectively. In
Figs.1 and 2, we represent the NSD $P(S)$ and the $\Delta_{3}$
statistics of the states with low spin-parity $J^\pi=6^+$ in U(5)
symmetry, respectively. The results at different sets of
parameters (in fact, different only in parameter $\beta$) are
marked with (a), (b), (c) and (d). In Figs.3(a)-(d) and 4(a)-(d),
we illustrate the results for the states $J^{\pi} = 6 ^{+}$ with
four values of parameter $\beta$ in O(6) symmetry. It has been
known that the classical limit of IBM corresponds to the system
with boson number $N \rightarrow \infty$. To show the finite boson
number effect, we have calculated the level statistics ($P(S)$ and
$\Delta_{3}(L)$) in each set of parameters for $N=25$, $N=70$ and
$N=200$. The results for these different $N$ are displayed in the
left, middle and right panel of the figures, respectively.
Meanwhile the Brody parameter $\omega$ of the level spacing
distribution\cite{Brod81} is also evaluated. The obtained results
are shown in the figures. These figures show that the results in
the two symmetries are quite similar to each other.

Comparing the results with the same boson number N but different
parameters, one can realize that, when the boson number $N$ has a
value not very large (e.g., 25, which is close to a realistic one
in nuclei), the statistics may show Poisson-type, GOE-type,
intermediate between Poisson-type and GOE-type, depending on the
values of the parameters. Comparing the results with different
boson numbers but in one set of parameters, one can know that when
$N\rightarrow\infty$, the statistics trends to be Poisson-type
independently of the interaction parameters. It indicates that the
finite boson number effect is prominent. Meanwhile, looking
through the Figs.~1-4 more cautiously, in particular the spectral
rigidity $\triangle_{3}(L)$ in Figs. 2(c) and 4(c), one may find
that for $N=25$ the results of the spectral rigidity lie outside
the lines for Poisson and GOE since the rigidity increases very
rapidly as the interval $L$ increases. If we return to the
Hamiltonian more carefully, we will find that the interaction of
the parent group is much stronger than that of the subgroup, which
may cause some big steps in the staircase function $N(E)$. These
big steps in different quanta of the parent group can not be
eliminated even after the energy levels are unfolded, and it will
cause some large level spacings in the NSD $P(S)$. As these
spacings are so large, they make most of the rest spacings less
than 1 since the average spacing is 1. Then appears the abnormal
NSD $P(S)$ statistics. For the spectral rigidity, we extract a
part of the unfolded staircase function $N(x)$ and illustrate it
in Fig.~5 with a rescaling, which involves a big level spacing
about 6. When $L$ increases from 5 to 10, it is clearly that the
local average least square deviation increases much more
drastically than only become twice. It should also be mentioned
that these abnormal statistics is in fact Poisson statistics. The
looking of departure from the Poisson-type statistics is because
the number of these large level spacings which affect strongly the
statistics is not very large(less than $N$). So they distribute
randomly in the NSD $P(S)$ and do not follow the statistical
principles. If we do not consider these large abnormal spacings,
the level spacings which are less than 1 will shift to following
Poisson statistics as the unfolded constant will decrease. The
spectral rigidity will not increase so fast as they are left out.

It should be mentioned that, in our calculations, all the
degenerate states are taken into consideration just as one single
state. That means, if the quantum numbers of some states differing
from others only in an additional quantum number, we take the
energy levels of these states as one single level when the energy
level statistics is carried out. On the contrary, if we regard
them as distinctive levels, the degeneracy causes so many zero
level spacings that the distribution is over-Poisson type. Fig.~6
displays the results for the states $J^{\pi}=6^{+}$ of the $N=25$
system with the degenerate states being taken into account
distinctively. Comparing Figs.~1-4. with Fig.~6, one can easily
recognize that the difference between the results with and without
the degeneracy being considered explicitly is apparent. In
practical calculation, nearly 1/4 levels of all are abandoned when
we have chosen just one level out of each set of the degenerate
states. It is obvious that such a manual selection of the levels
introduces a finite symmetry breaking to lift the degeneracy.

It has been known that the degeneracy results from the absence of
sufficient quantum numbers due to symmetry. In previous numerical
calculations where the transitions from one dynamical symmetry to
the other were investigated, since the symmetries have been
broken, the degeneracy is then broken, such a problem seems do not
exist. However, when we analyze the statistics in the dynamical
symmetries, we have to handle the problem since the additional
quantum number $K$ in Eqs.(4-6) can be quite large if the boson
number $N$ is large. In the previous investigations on the
statistics of the energy levels in SU(3)
symmetry\cite{PV90,PVD92}, Paar and collaborators discussed the
case of $J^\pi=0^+$, where the additional quantum number $K$ takes
only one value $K=0$, and also the case of $J^{\pi} \ge 2^+ $,
where $K$ could have more than one values. Such an additional
quantum number $K$ may be viewed as a result of a hidden
symmetry\cite{Ku97} since the states with the same angular
momentum but different $K$ are degenerate. The calculated results
showed that, for $N=20$ which is not very large, the statistics of
the states $J^\pi=0^+$ is close to GOE-type. As the boson number
$N$ increases, the statistics gets close to Poisson-type. For the
states $J^\pi=2^+$, if the $K$ is fixed to a certain number, the
energy level statistics is closed to GOE-type. In the present
work, we also analyze the case of SU(3) symmetry with different
boson numbers. In the analysis we select only one level from each
set of degenerate states to establish the level set for
statistics, which is just the same as that taken for the U(5) and
O(6) symmetries, and is equivalent to that with $K\equiv{0}$ in
Paar's work. The obtained results for the states of $J^\pi=6^+$ in
the systems with boson number $N=25$, 70 and 200 are illustrated
in Fig.~7. The calculated results show that the trend of
statistics from GOE-type to Poisson-type as $N$ increases is
clear, which coincides with the result of Paar and
collaborators\cite{PVD92}.

For the sake of a self consistent discussion, we also show
explicitly the energy level statistics of the states $J^\pi=0^+$
for which the degeneracy caused by hidden symmetry does not exist.
Since numerical results show that the statistics of the O(6)
symmetry is quite close to that of the U(5) symmetry, we
represented then only the results for the $P(S)$ and
$\Delta_{3}(L)$ statistics in U(5) dynamical symmetry in Figs.~8
and 9, respectively. It is clear that the energy level statistics
is quite similar to that in Figs.~1 and 2, as we have discussed
before. It manifests that our removal by hand of all but one of
the degenerate states is equivalent to that without
multi-degeneracy .

As mentioned above, in order to show how the manually introduced
symmetry breaking affects the statistics, we also calculate the
statistics with distinctive degenerate states. The results for the
case with $N=25$ are shown in Fig.~6. Comparing Fig.~6 with
Figs.~1-4 in the case of the same parameters, one can easily infer
that the manually introduced symmetry breaking makes the
statistics from over-Poisson type to GOE type. The results are
quite consistent with Paar and collaborators' work\cite{PV90},
where they introduce an additional term which breaks the $K$
quantum number but conserves SU(3) dynamical symmetry. Their
results show that increasing the strength of the $K$ breaking term
makes the statistics change continuously from over-Poisson type to
GOE type. One thing we need then to point out here is that in
practical calculation, the hidden symmetry is completely broken
not in the case that the degeneracy resulting from the existence
of additional quantum number is removed(types of the interaction),
but in the case that the strength breaking the symmetry reaches a
certain value(strength of the interaction). This might interpret
why in Alhassid and collaborates' work\cite{AN92}, the statistics
near the dynamical symmetries is in an ``overintegral" situation
with negative Brody parameter $\omega$. When the Hamiltonian they
use become very close to the one in the dynamical symmetry, for
instance, when $c_0 = 0$ and $\chi = -0.01$ in the self-consistent
$Q$ formalism\cite{WC82} near the O(6) dynamical symmetry, the
broken strength is too weak to break the hidden symmetry resulting
from the missing labels though the degeneracy does not exist. Then
the question comes out that which way to determine the level set
can better describe the statistics of the realistic nucleus in the
dynamical symmetry. In the present O(5)$\supset$ O(3) reduction,
the degeneracy due to the hidden symmetry is distinguished by the
manually introduced additional quantum number but no interaction
is involved to link the states with different additional quantum
number $K$. Therefore the degenerate states with different
additional quantum numbers are in fact statistically uncorrelated.
When we calculate the fluctuations of the energy levels, such a
large amount of the statistically uncorrelated states should be
removed. Otherwise, a mixed ensemble (with different good quantum
number $K$) is taken into consideration and the over-Poisson type
distribution would be obtained (because nearly 1/4 of the spacings
of all are zero in practical calculations). In this point of view,
the ``overintegral" situation in Ref.\cite{AN92} may arise from
that some statistically uncorrelated energy levels were taken into
consideration.

In Fig.~7, we display the results with only one set of parameters
$\gamma=0.01$, $\delta=-0.7$ (in arbitrary unit), because from
Eq.(9), one can know that different values of parameters $\gamma$
and $\delta$ cause only a linear transformation of the energies.
Then it does not affect the statistics. Analogously, one may get a
conclusion from Eqs.(7) and (8) that changing the parameter should
not affect the statistics in U(5) and O(6) symmetries since the
type of the interaction and the structure of the energy levels
remain the same. However, recalling Figs.~1-4, one can realize
that, if the boson number is not very large(e.g. $N$=25 ), the
statistics in U(5) or O(6) symmetry depends obviously on the
absolute value of the parameter $\beta$. This indicates that the
relative strengths of the interactions with different symmetries
also affect the statistics. If we take the results more carefully,
we will find that the increase of the relative value $\beta$ (in
fact $\beta / \alpha$, $\beta / \eta$) makes the statistics in
realistic case $(N=25)$ change gradually from the Poisson-type to
GOE-type. In order to show the dependence of the statistics on the
interaction strengths more obviously, we calculate the Brody
parameter $\omega$ of the level spacing distribution in a wide
range of parameter $\beta$ in U(5) and O(6) symmetries,
respectively. The obtained results are given in Figs.~10(a),(b),
respectively. The figures show that the statistics varies from
Poisson-type to GOE-type, and further to Poisson-type again with
respect to the increasing of $\beta / \alpha$, $\beta / \eta $.
Recalling Eqs.(7) and (8), we can realize that for a small value
of $\beta / \alpha$, $\beta / \eta$, the interaction with the O(5)
symmetry is only a perturbation on the U(5), O(6) symmetries, the
quantum system is then approximately regular. While the ratio
increases, the interaction strength of the O(5) becomes comparable
to the strength of the parent group U(5) or O(6), the statistics
appears in GOE-type. It indicates that, when the strengths of the
interactions with different symmetries are comparable and compete
with each other in one quantum system, chaos may come out. This
mechanism of onset of chaos can also be seen in Alhassid and
collaborates' work of investigating the broken pairs in
nuclei\cite{AV92}, where when the Coriolis interaction is
comparable to the pairing interaction, the degree of chaoticity
seems to be maximal. As the ratio of the interactions changes
further and becomes so large that the interaction of the parent
group plays only a role as a ``perturbation", the quantum system
recovers approximately regular.

It is worth mentioning that the above results are quite similar to
the well-know case of the hydrogen atom in a uniform magnetic
field\cite{FW89}. The Sturm-Coulomb problem is an integrable one
since it holds O(4) symmetry. When one puts the atom into a
magnetic field, the O(4) symmetry is broken and reduced to the
O(2) symmetry. The problem becomes then nonintegrable. The chaos
arises and is being obvious when the energy or the magnetic
strength (connected to O(2) strength) increases. The present
U(5)$\supset$ O(5) and O(6)$\supset$ O(5) reductions are analogous
to the O(4)$\supset$ O(2). The onset of chaos in the U(5) and O(6)
symmetry is a direct result against the corresponding increase of
the interaction strength of O(5) symmetry since the symmetry is,
in fact, broken.

Aside from the above analogy, another problem might have some
relation with the above results. In the geometric analysis of IBM,
the subgroup O(5) in the dynamical algebra U(6) corresponds only
to the kinetic part of the Hamitonian\cite{Hat82}, which can be
written as
$$\displaylines{\hspace*{1cm}
T_{\gamma}=p_{\gamma}^2+\frac{1}{4}\sum_{m=1}^{3}\frac{L_{m}^{2}}{\sin^{2}(\gamma-2\pi
m/3)} \,  \hfill{(19)} \cr }
$$
in the classical limit. As a result, the O(5) symmetric term can
be viewed as the kinetic interaction that does not affect the
$\beta-\gamma$ dependence of the potential surface and contains
only the collective motion of the nucleus according to Leviatan
and collaborates' work\cite{Levi87}. Indeed, the interaction
strength of the subgroups in the system, such as that of the O(5)
symmetric one in the IBM, can be viewed as parts of the dynamical
origin of the chaotic behavior, or more concretely, the GOE
fluctuations in nuclei. Just as Bohigas pointed out\cite{Boh88}
:``In our opinion, the static nuclear mean field is too regular to
be held responsible, and chaos must be caused by the residual
interaction." In this point of view, the onset of chaos in nucleus
results not only from the symmetry breaking of the potential(mean
field), but also from the increasing of the interaction strength
of the subgroup(residual interaction), which is competing with the
parent group.

Furthermore, Eq.(7) shows that the energy of the states in U(5)
symmetry depends not only on parameter $\beta$, but also on the
other parameters, such as $\alpha$ and $\varepsilon$. Considering
the geometric model correspondence of the IBM, one knows that U(5)
symmetry corresponds to an inharmonic vibration with frequency
$\hbar\omega=\varepsilon+(n_{d}+4)\alpha$. It is obvious that,
with the increasing of the $d$-boson number, if $\alpha>0$, the
vibration frequency increases; if $\alpha<0$, the vibration
frequency decreases. It has recently been shown that the U(5)
symmetry with parameter $\alpha<0$ can describe the collective
backbending of high spin states well\cite{Long97}. For a given set
of parameters with $\alpha<0$, there exists a critical angular
momentum
$$\displaylines{\hspace*{1cm}
L_{c}\approx-\frac{2(\varepsilon+4\alpha)}{\alpha}-2N \, ,
\hfill{(20)} \cr }
$$
where $N$ is the total boson number. As the angular momentum $L
\geq L_c$, the yrast states are no longer the inharmonic
vibrational states, but the rotational ones with $n_d=N$. For a
system with $N$=25 and parameters $\varepsilon=3.53 $,
$\alpha=-0.101$(in arbitrary unit), in which $L_c$=12, we analyze
the energy level statistics of the states $L=6(<L_c)$,
$L=12(=L_c)$ and $L=20(>L_c)$. The obtained results are
illustrated in Fig.~11. It is apparent that two maxima appear in
the nearest neighbor level spacing distribution $P(S)$. Such a
behavior is quite different from the fluctuation properties in
other cases. In theoretical point of view, the term
$n_{d}\varepsilon$ enlarges the level spacing, whereas the term
$n_d(n_{d}+4)\alpha$ with $\alpha<0$ compresses the level spacing.
The simultaneous appearance of these two effects induces a
competition which makes the energy of the states in the ground
state band of the U(5) symmetry
$$\displaylines{\hspace*{1cm}
E_{gsb}(n_d)=(\alpha+\beta+4\gamma){n_d}^{2}+(\varepsilon +
4\alpha+3\beta+2\gamma)n_{d} \,  \hfill{(21)} \cr }
$$
not increase with respect to the increasing of the d-boson number
monotonously. Then a maximal d-boson number limit $n^{(m)}_{d}$
for the energy to increase correspondingly exists
$$\displaylines{\hspace*{1cm}
n^{(m)}_{d}=\frac{\varepsilon+4\alpha+3\beta+2\gamma}{-2(\alpha+
\beta+4\gamma)}\approx\frac{\varepsilon+4\alpha}{-2\alpha} \, .
\hfill{(22)} \cr }
$$
Considering the property of the parabola $E_{gsb}(n_d)$, one can
know that there exists also a boson number
$n_{d}^{(u)}=2n_{d}^{(m)}-N$, with which the energy of the system
equates to that with $n_{d}=N$, and those with
$n_{d}\in(n^{(u)}_d, n^{(m)}_d)$ are larger than those of the
states with the same angular momentum but $n_d=N$. It means that
the states in the intrinsic ground state band with d-boson number
$n_{d}\in(n^{(u)}_d, N)$ are no longer the yrast states. Then the
structure of the yrast band changes from the U(5) states to the
rotational states with d-boson number $n_{d}=N$ and the energy of
the states in the yrast band changes in the way $L(L+1)$. It
implies that a phase transition of collective motion mode may
happen as the angular momentum reaches the critical value
$(L_{c}=2n_{d}^{(u)})$ in Eq.(20).

We can also speculate it more clearly in the thermodynamical
analysis. If we use the coherent state formalism\cite{GK80,DSI80}
of the IBM, the energy functional for the U(5) Hamiltonian is
given by
$$\displaylines{\hspace*{1cm}
E(N, f_{1}, f_{2}; \beta) = E_{0} + f_{1} N
\frac{\beta^{2}}{1+\beta^{2}} + f_{2}
N(N-1)\frac{\beta^{4}}{{(1+\beta^2)}^{2}}. \hfill{(23)} \cr }
$$
The parameter of the first term $f_{1}$ depends on the parameter $
\varepsilon, \beta, \gamma$ used in Eqs.(1) and (7), while the
parameter $f_2$ of the second term depends only on the parameter
$\alpha$. It is definite that if the interaction shown in the
second term is attractive, which corresponds to $\alpha<0$ and
$f_{2}<0$, there exist two minima for the energy functional in
Eq.~(23). One is $E_0$ arising from $\beta = 0$. Another is $E_0 +
f_1 N + f_2 N(N-1)$ corresponding to $\beta \rightarrow \infty$.
In the general framework of the Landau theory of phase
transitions\cite{LL01}, it has been well known that, for a system
with thermodynamic potential $\Phi(P, T; \xi)$ that depends on
external parameters (pressure P and temperature T) and the order
parameter $\xi$, the first-order transitions are characterized by
a singularity in the specific heat $C_{p} =
-T\partial^{2}\Phi/\partial T^{2}$, which gives a nonzero latent
heat. Meanwhile, the optimal order parameter $\xi_0$ to minimize
the functional jumps {\it discontinuously} from one value to
another. It has been well known that the deformation parameter
$\beta$ is the quantity to characterize the shape of a nucleus, it
can then be taken as the order parameter to identify the nuclear
shape transition. Then, instead of the thermodynamic potential
$\Phi(P, T; \xi)$, the $E(N, f_{1}, f_{2}; \beta)$ in Eq.~(23) can
be taken as the ``thermodynamic potential"\cite{CJ98,JCCHLW02}.
Analogous to the above mentioned general case, the {\it
discontinuous} jump of the minima of the energy functional $E(N,
f_1, f_2, \beta)$ against the order parameter $\beta$ (from $0$ to
$\infty$) indicates that the shape phase transition discussed here
is in first-order. This is in agreement with the results found for
the vibron model\cite{I78,IL95} with random interactions, which
exhibits a first order phase transition between a ground state
with $n_p=0$ and $n_p=N$\cite{BF02}. It is also remarkable that
the rotation after the shape phase transition is the one with
variable-momentum since the stable state corresponds to the
asymptotic process $\beta \rightarrow \infty$.

It has been known that, as a general rule, phase coexistence can
be found in first-order phase transitions\cite{LL01}. In our
present energy level statistics analysis, the appearance of
different shapes makes the ensemble (with the same angular
momentum $L$) in the U(5) symmetry with $\alpha < 0$ involves in
fact two sequences, one of which is in vibration, another one is
in rotation. When we unfold the above spectrum $\{E_{i}(L)\}$, the
two sequences are normalized with an unique total average spacing,
and their maxima of the spacing distributions do not appear at the
same value of $S$, As a consequence, two maxima emerge. It should
be noted that there are only few overlaps between the two
different sequences, otherwise the overlap of the two sequences
will change the nearest neighbor distribution. On the other hand,
recently two new symmetries denoted by E(5)\cite{Iac00} and
X(5)\cite{Iac01,ZBCJ02} have been proposed to describe the spectra
of nuclei at or around the critical point of shape transitions
from vibration to $\gamma$-unstable rotation (second-order) and
those from vibration to axial rotation (first-order),
respectively. And the X(5) symmetry describes the nuclei at or
around the critical point of the first-order shape phase
transition which involves shape coexistence\cite{IZC98,Zamf02}.
For comparison, We also analyze the energy level statistics of the
states with E(5) and X(5) symmetries. Our results show that the
NSD $P(S)$ of X(5) symmetry exhibits also two maxima(the full
result will be published elsewhere\cite{SJL02}). Such a similarity
in the NSD might provides further evidence that the collective
backbending is a manifestation of shape phase transition in first
order and involving shape coexistence.

To manifest the constituent of the level ensemble, we evaluate the
rigidity $\Delta_{3}(L)$ more meticulously, too. The result is
illustrated in Fig.~12. Fig.~12 shows that the $\Delta_{3}(L)$ of
the spectrum with collective backbending exhibits a drastic
fluctuation. It has been known that the variance of the
$\Delta_{3}$ statistics $(\langle \Delta _3 ^2 \rangle - \langle
\Delta _3 \rangle ^2)$ is connected with the 3- and 4-level
correlations, which is expected to be very small in the
past\cite{meta67}. The conflicts between the obtained results and
general behavior indicts that the system undergoing a shape phase
transition might exhibit a much strong fluctuations than the usual
systems discussed in the past.

For comparison, we evaluate the statistics of the system with
$N=70$, 200 for $\alpha<0$, too. The calculated results for $L=6$
states are represented in Fig.~13. One can know from Eq.(20) that,
for $N$=70 or 200, $L_{c}<0$ if $\varepsilon$ and $\alpha$
maintain their values as the same as those for $N$=25, the
competition mentioned above does not play any role for all the
states. Then the two maxima in the NSD statistics and the drastic
fluctuation in the $\Delta_{3}$ statistics no longer exist. The
appearance of the Poisson-type distribution indicates that only
the rotational mode plays important role in the system. Comparing
the result with $N$=25 and those with $N$=70, 200, one can reach a
conclusion that the appearance of the collective backbending is a
signal of phase transition from a vibration to a rotation.
Meanwhile the emergence of two maxima in the $P(S)$ distribution
may be a characteristic of the shape phase transition.

\section{SUMMARY AND REMARKS}

In summary, we have analyzed the energy level statistics of the
U(5), O(6) and SU(3) dynamical symmetries in the interacting boson
model(IBM) in this paper. In the analysis, the degeneracy
resulting from the additional quantum number was eliminated
manually, i.e. we took only one level from each set of degenerate
states. The calculated results indicate that the finite boson
number $N$ effect is prominent. If $N$ takes a value not very
large(e.g., 25, which is close to a realistic one), the statistics
depends strongly on the interaction strength of the subgroup O(5).
While $N \rightarrow \infty$, they all trend to be Poisson-type.
We would like then to mention that the interaction of the subgroup
O(5) which only possess the collective motions of the nucleus can
be viewed as the dynamical origin of chaos in nucleus and the
symmetry paradigm deserves more careful consideration. In fact,
exceptions to the symmetry paradigm have been found for many years
(see, for example, Ref.\cite{CC89}).

In this paper, we also analyzed the level statistics of the states
holding the collective backbending in high spin states. We found
that the nearest neighbor level spacing distribution $P(S)$ of the
states with the collective backbending involved two maxima and the
$\Delta_{3}$ statistics exhibited a fierce fluctuation, which were
drastically different from the general properties in each symmetry
of the IBM at usual situation. It indicates that the spectrum
involves a shift between two modes of collective motions. It
provides then a clue that the collective backbending is a
characteristic of shape phase transition. Furthermore, looking
through all the process, we can suggest that the statistics of the
system can result from not only the form of interaction (the
Hamiltonian or perturbation) but also the interaction strength.

\vspace*{8mm}

This work is supported by the National Natural Science Foundation
of China under the contract No. 19875001, 10075002, and 10135030
and the Major State Basic Research Development Program under
contract No.G2000077400.
One of the author (J. S.) thanks also the
support from the Taizhao Foundation at Peking University. Another
author (Y.X. L.) thanks the support by the Foundation for
University Key Teacher by the Ministry of Education, China, too.

\newpage

\parindent=0pt

{\bf Figures and Their Captions:}

\begin{description}
\item{Fig.~1.} Comparison of nearest-neighbor spacing distribution
$P(S)$ of the states $J^\pi=6^+$ in U(5) symmetry with four sets
of parameters: (a) for $\varepsilon =1.76$, $\alpha=0.1$,
$\beta=0.02$, $\gamma=0.001$, (b) for $\varepsilon=1.76$,
$\alpha=0.1$, $\beta=0.01$, $\gamma=0.001$, (c) for
$\varepsilon=1.76$, $\alpha=0.1$, $\beta=0.006$, $\gamma=0.001$,
(d) for $\varepsilon=1.76$, $\alpha=0.1$, $\beta=-0.01$,
$\gamma=0.001$. In all figures, the solid lines and dashed lines
describe the GOE and Poisson statistics, respectively.

\item{Fig.~2.} The same as Fig.~1 but for the spectral rigidity
$\Delta_3(L)$.

\item{Fig.~3.} Comparison of nearest-neighbor spacing distribution
$P(S)$ of the states $J^\pi=6^+$ in O(6) symmetry with four sets
of parameters: (a) for $\eta=-0.5$, $\beta=0.15$, $\gamma=0.001$,
(b) for $\eta=-0.5$, $\beta=0.10$, $\gamma=0.001$, (c) for
$\eta=-0.5$, $\beta=0.05$, $\gamma=0.001$, (d) for $\eta=-0.5$,
$\beta=-0.10$, $\gamma=0.001$.

\item{Fig.~4.} The same as Fig.~3 but for the spectral rigidity
$\Delta _3 (L)$.

\item{Fig.~5.} Comparison of the local average least square
deviation to a part of staircase function $N(E)$ (solid steps).
The levels analyzed are the $J^{\pi}=6^{+}$ levels of the same
U(5) Hamiltonian as Fig.~1(c) with boson number $N=25$. The dash
lines and dashed-dotted lines describe the local average least
square lines as the interval $L$ equates 5 and 10, respectively.

\item{Fig.~6.} Energy level statistics for U(5) and O(6) symmetries
with distinctive degenerate states caused by additional quantum
number in the states $J^{\pi}= 6^+$ when boson number $N=25$. (a)
For U(5) symmetry with parameters $\varepsilon=1.76$,
$\alpha=0.1$, $\beta=0.01$, $\gamma=0.001$. (b) For O(6) symmetry
with parameters $\eta=-0.5$, $\beta=0.15$, $\gamma=0.001$.

\item{Fig.~7.} Energy level statistics for the $J^{\pi}=6^{+}$ states
in SU(3) symmetry with different number of bosons.

\item{Fig.~8.} The same as Fig.~1 but for angular momentum
and parity $J^{\pi}= 0^+$.

\item{Fig.~9.} The same as Fig.~2 but for angular momentum
and parity $J^{\pi}= 0^+$.

\item{Fig.~10.} Quantum measures of chaos in different
interaction strength of O(5) symmetry:  (a) Brody parameter
$\omega$ versus $\beta/\alpha$ for $\varepsilon =1.76$,
$\alpha=0.1$, $\gamma=0.001$ in U(5) symmetry, (b) Brody parameter
$\omega$ versus $\beta/\eta$ for $\eta=-0.5$, $\gamma=0.001$ in
O(6) symmetry.

\item{Fig.~11.} Energy level statistics of the states with boson number
$N=25$, angular momentum and parity $J^\pi=6^+$($L<L_c$),
$12^+$($L=L_c$), $20^+$($L=L_c$) in U(5) symmetry when it has
collective backbending at high spin states (The parameters are
$\varepsilon=3.51$, $\alpha=-0.101$, $\beta=0.01$,
$\gamma=0.001$).

\item{Fig.~12.} Comparison of $\Delta_{3}$ statistics of the states
in U(5) symmetry with boson number $N=25$ with collective
backbending and without collective backbending: (a) for
$\varepsilon=3.51$, $\alpha=-0.101$, $\beta=0.01$, $\gamma=0.001$,
$J^\pi=6^+$($L<L_c$), (b) the same parameters with (a) but for
$J^\pi=12^+$($L=L_c$), (c) the same parameters with (a) but for
$J^\pi=20^+$($L=L_c$), (d) for $\varepsilon =1.76$, $\alpha=0.1$,
$\beta=0.02$, $\gamma=0.001$, $J^\pi=6^+$.

\item{Fig.~13.} The same as Fig.~11 but with boson number $N=70$
and $200$ for the $J^{\pi} = 6 ^{+}$ states.
\end{description}

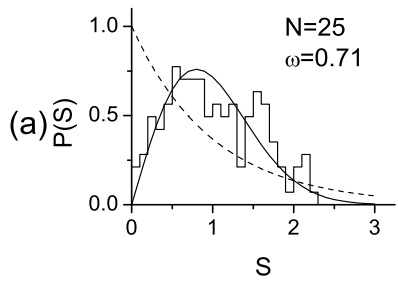
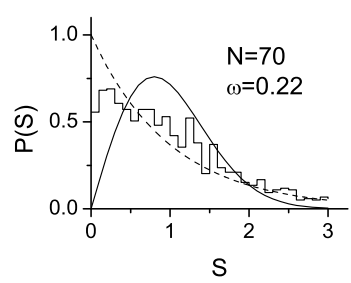
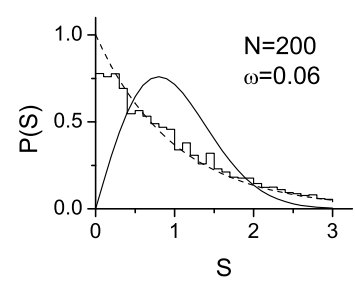
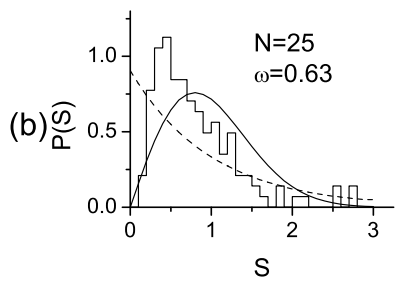
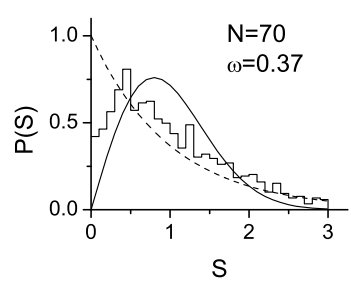
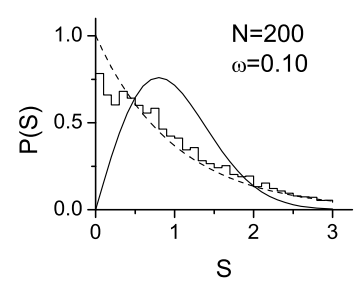
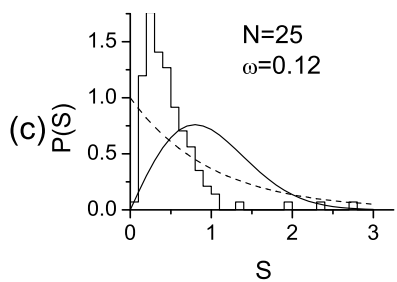
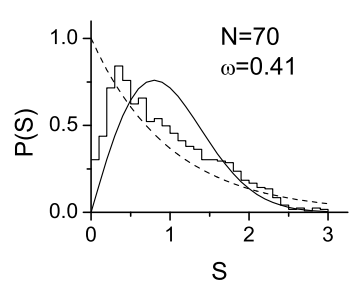
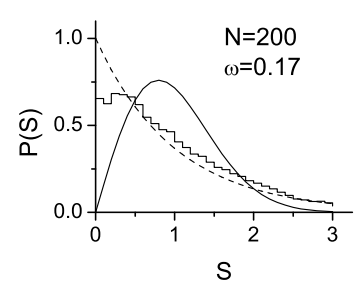
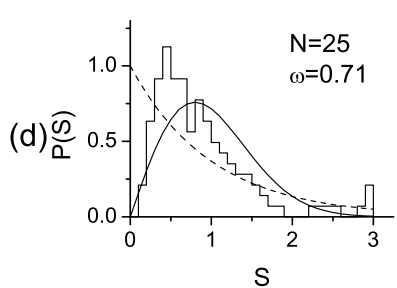
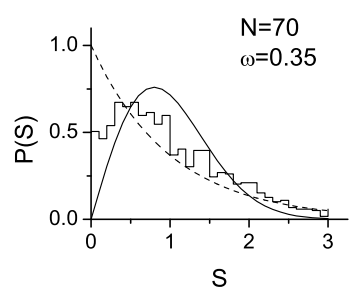
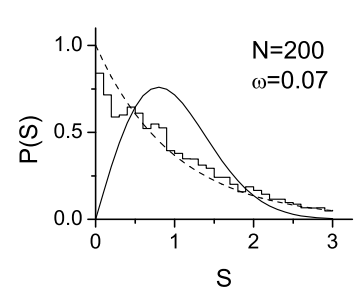
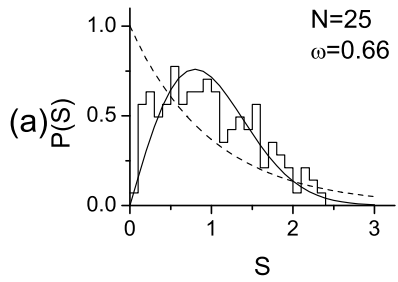
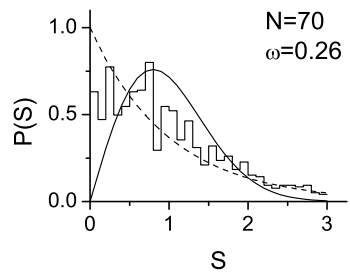
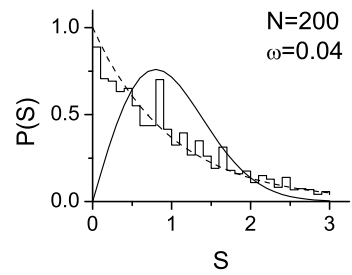
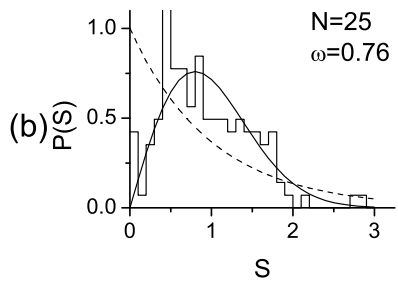
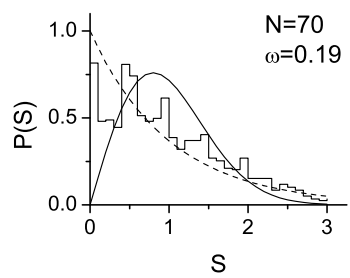
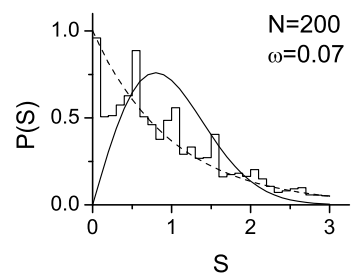
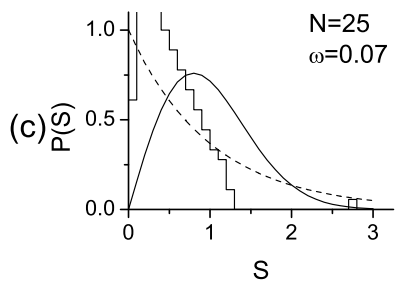
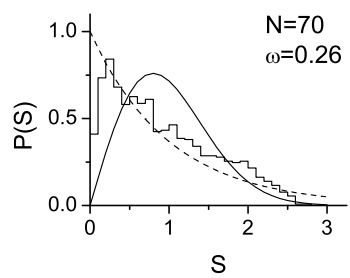
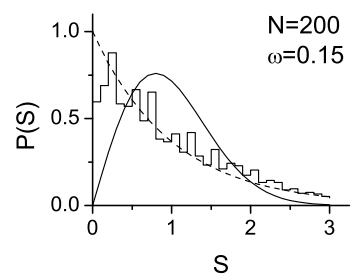
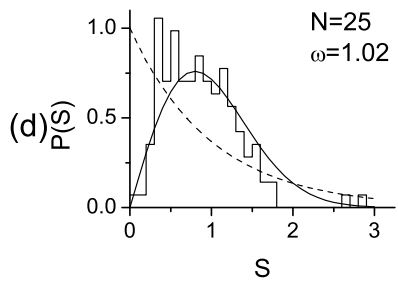
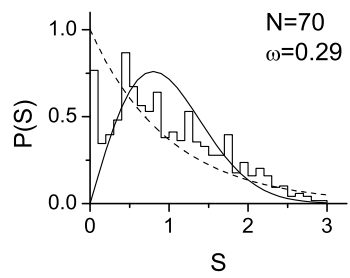
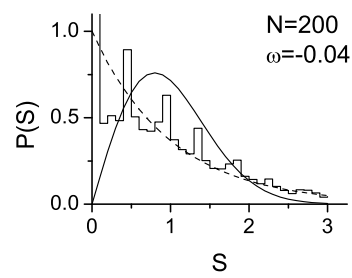
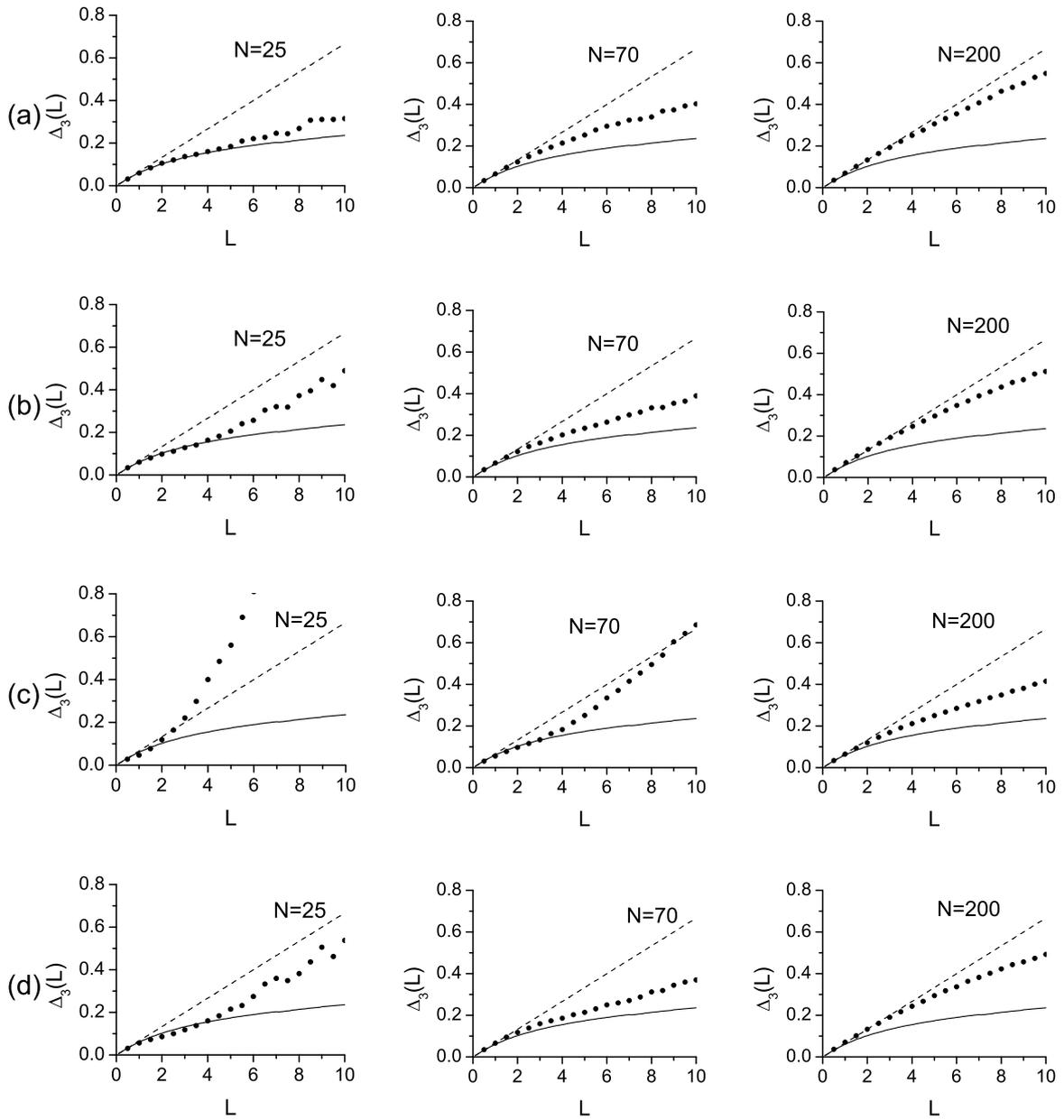
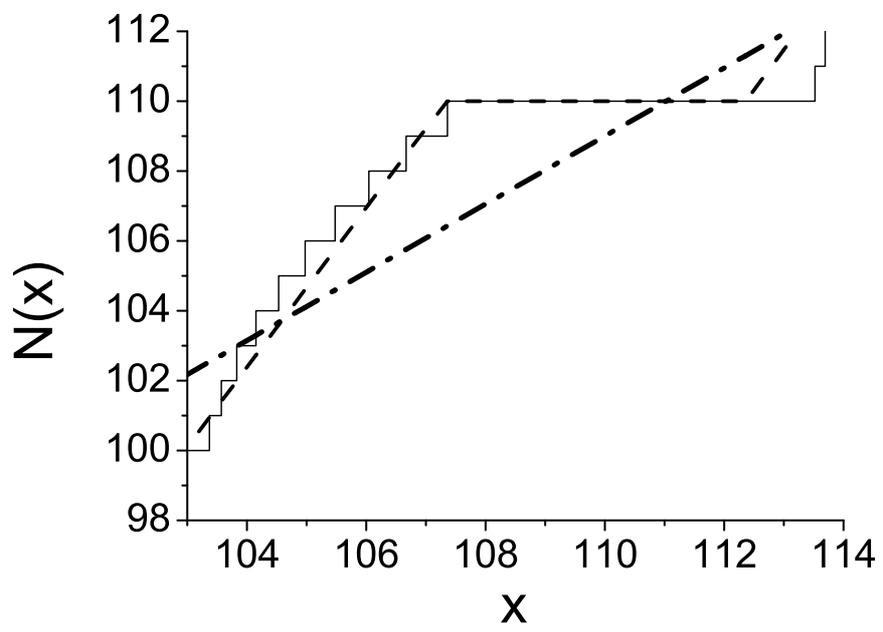
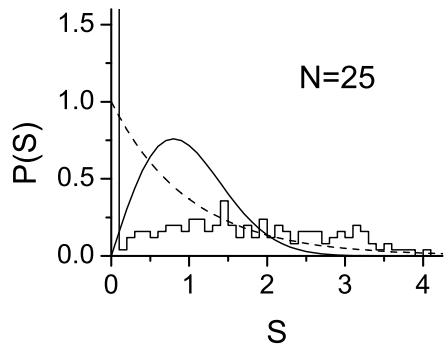
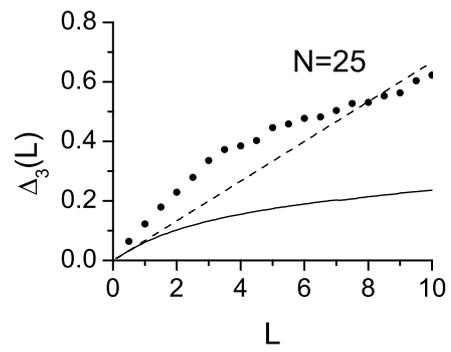
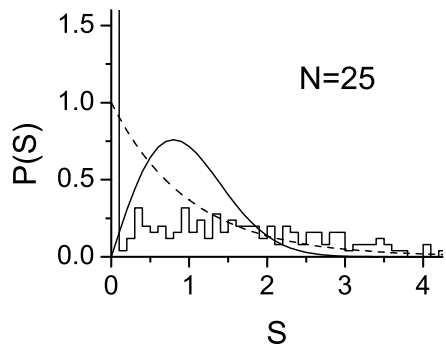
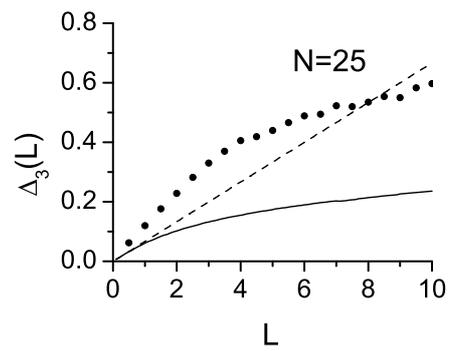
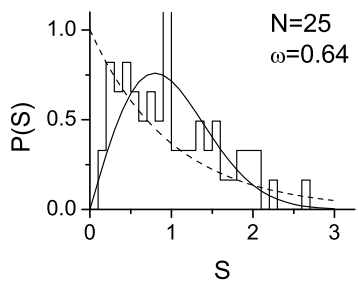
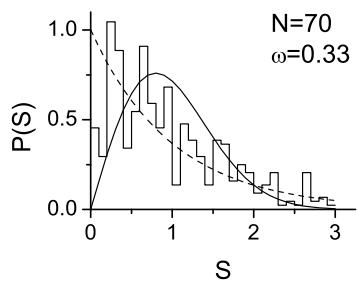
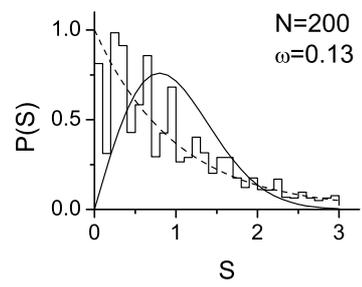
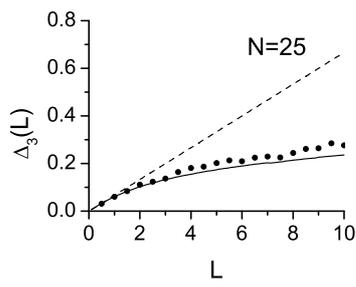
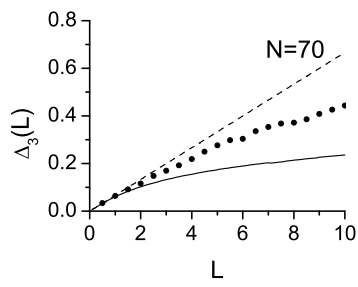
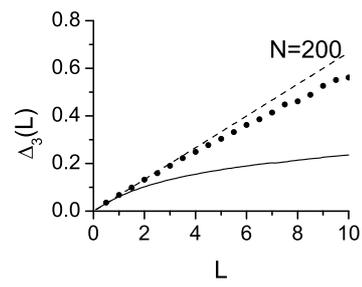
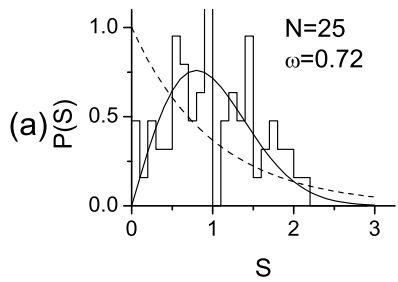
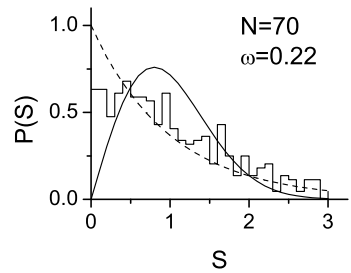
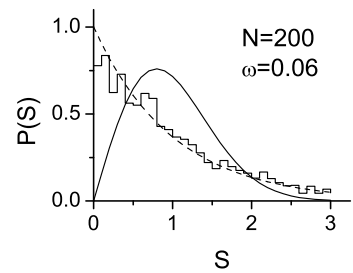
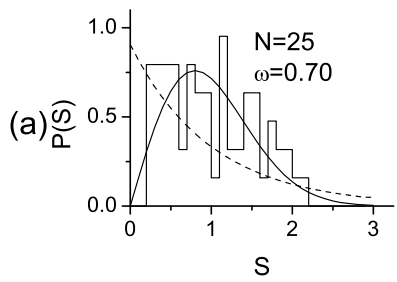
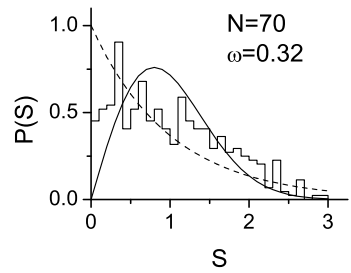
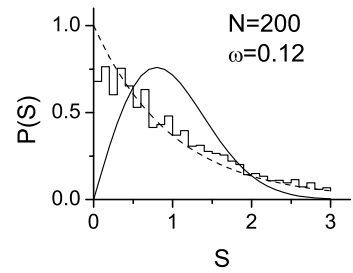
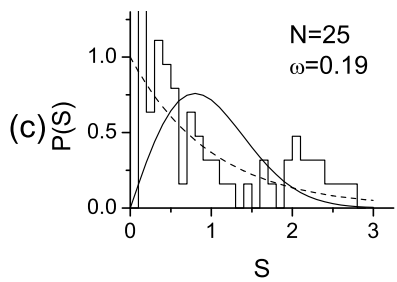
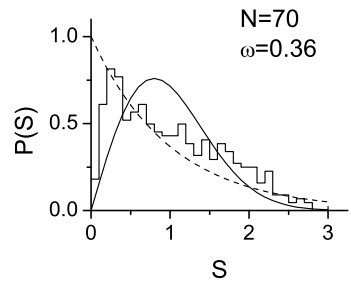
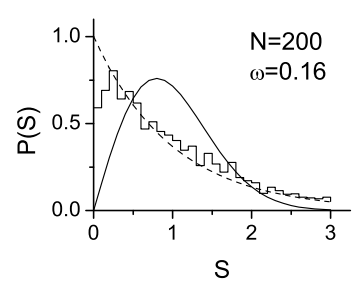
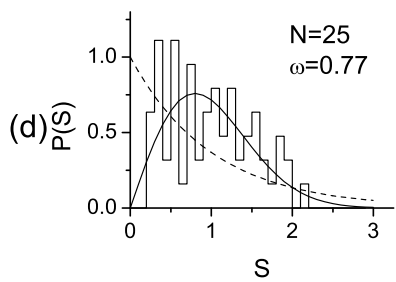
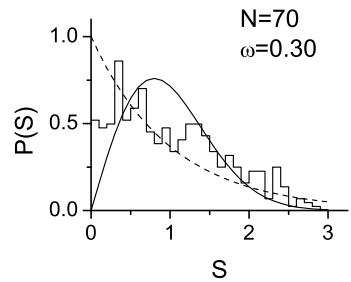
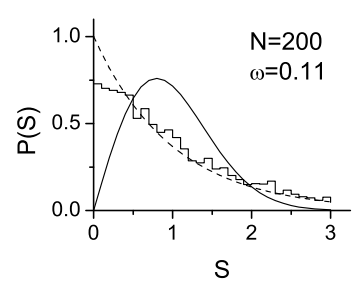
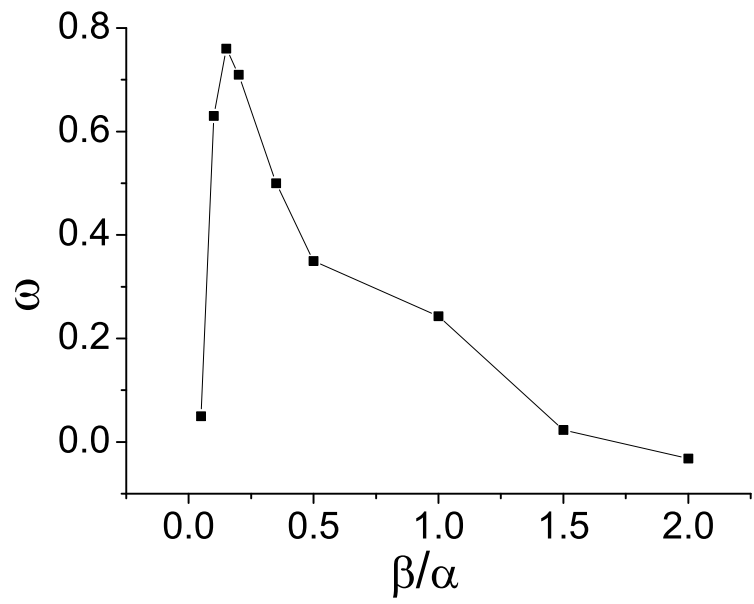
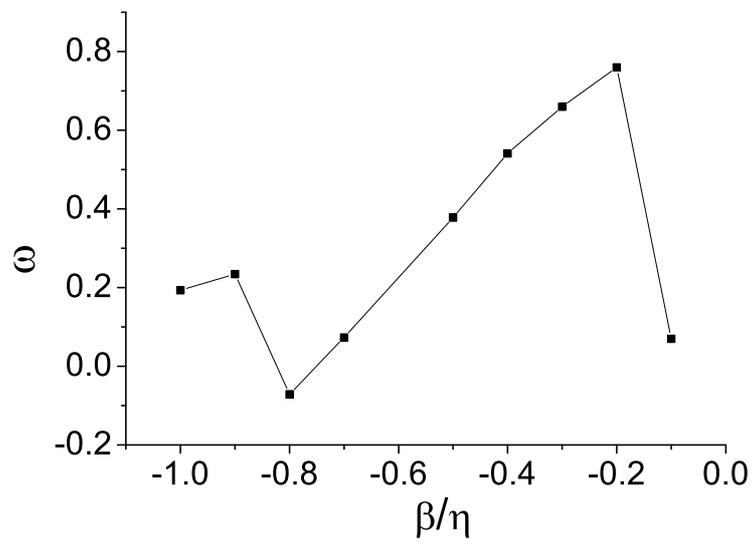
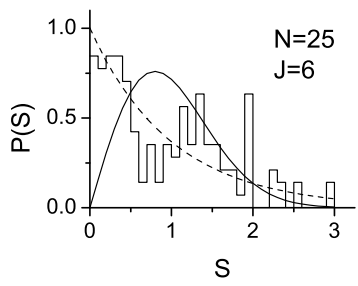
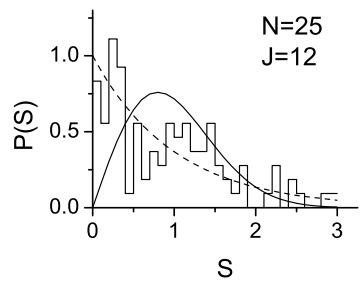
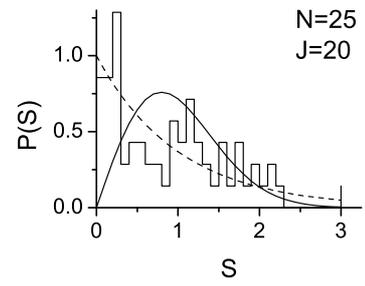
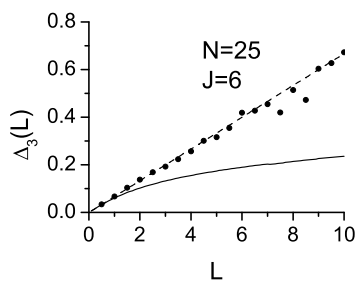
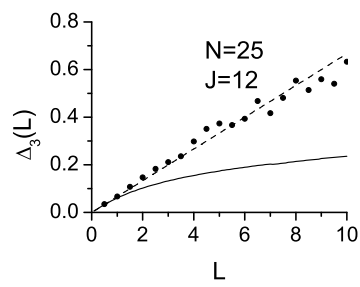
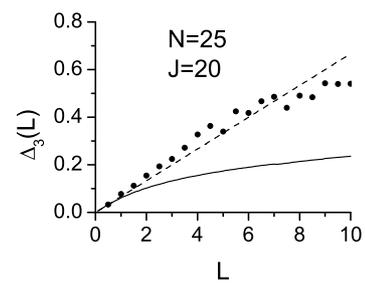
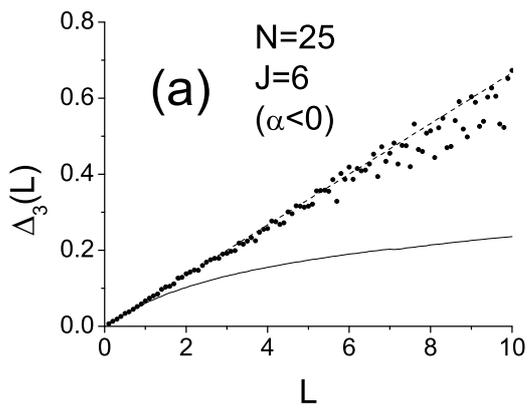
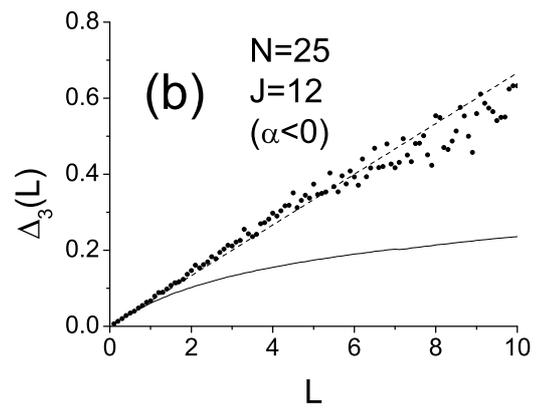
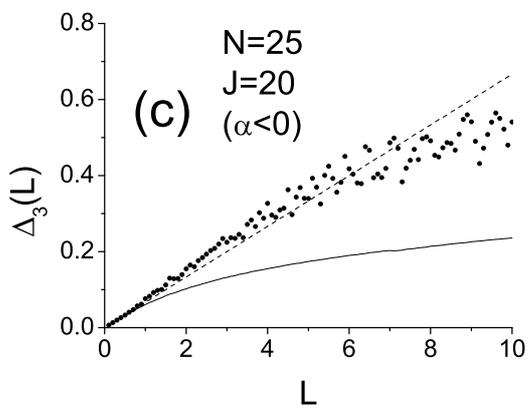
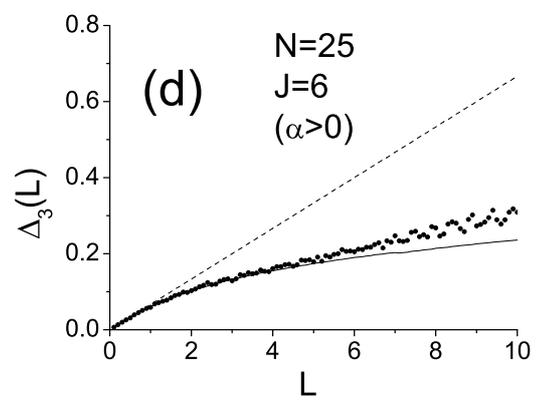
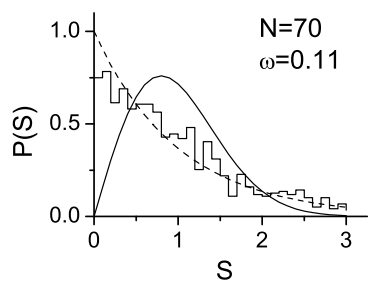
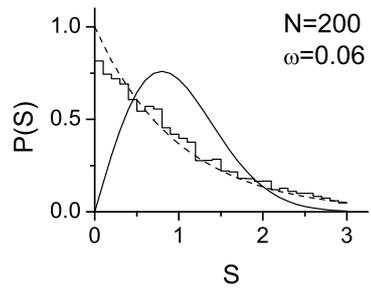
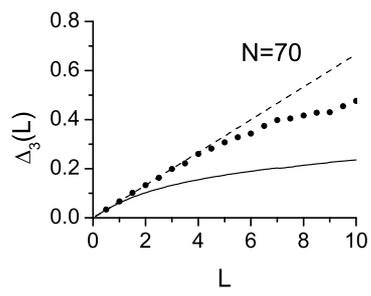
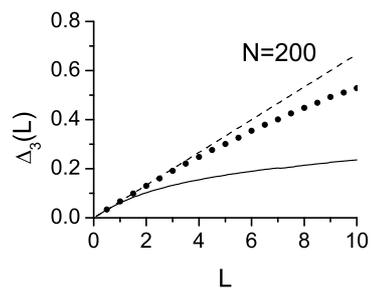
\end{document}